\documentclass[aps,showpacs,prb,floatfix]{revtex4}%
\usepackage{graphicx}
\usepackage{amsmath}
\usepackage{amsfonts}%
\usepackage{amssymb}
\begin{document}
\title{Non-periodic pseudo-random numbers used in Monte Carlo calculations}
\author{Gaston E. Barberis}
\date{\today }

\begin{abstract}
The generation of pseudo-random numbers is one of the interesting problems in
Monte Carlo simulations, mostly because the common computer generators produce
periodic numbers. We used simple pseudo-random numbers generated with the
simplest chaotic system, the logistic map, with excellent results. The numbers
generated in this way are non-periodic, which we demonstrated for 10$^{13}$
numbers, and they are obtained in a deterministic way, which allows to repeat
systematically any calculation. The Monte Carlo calculations are the ideal
field to apply these numbers, and we did it for simple and more elaborated
cases. Chemistry and Information Technology use this kind of simulations, and
the application of this numbers to Quantum Monte Carlo and Cryptography is
immediate. I present here the techniques to calculate, analyze and use these
pseudo-random numbers, show that they lack periodicity up to 10$^{13}$ numbers
and that they are not correlated.

\end{abstract}
\maketitle

\address{Instituto de F\'{\i}sica ''Gleb Wataghin'', UNICAMP, 13083-970 Campinas, (SP)\\
Brazil}

%main text

\section{Introduction}

Monte Carlo calculation is a very efficient method to simulate complex
physical systems. It is based in the generation of random numbers, and the
main applications of this method in physics are statistical and quantum
systems. Generally, instead of true random numbers, the researchers use
pseudo-random numbers, generated by computers. This has an important advantage
over the use of real random numbers: every calculation can be repeated several
times, if necessary, and the pseudo-random numbers generated repeatedly, in
order to check previous results. The main problem with those numbers is the
periodicity that appears near one billion generated numbers, together with
undesirable correlations, linear and no linear, that appear in the sets. Here
we present the use of the simplest equation where chaos occurs, the logistic
map.[1] \emph{\ }Using it, it is possible to generate random numbers in the
(0,1) interval, with a particular distribution, and using the Ulam-Neumann
transformation [2-5] they transform in a set of uniform deviates in that
interval. This paper is dedicated especially to the study of the pseudo-random
numbers generated with the logistic map, their periodicity and correlation.
The pseudo-random numbers generated with it are not periodic, allowing us to
create sequences as big as necessary to simulate statistical systems. The
random generated numbers, as described above, allow us a wide field of
applications. The obvious among them is also the simplest magnetic system
simulation, the 2D Ising ferromagnet. As this system was exactly solved by
Onsager, it allows to compare the exact results, the critical exponents and
properties with the simulations, so we use it as an example of application.
Other, we selected the chaotic cryptography as an exercise, using a simplified
ASCII symbols set. We applied the numbers to solve condensed matter systems,
as colossal magneto-resistance materials, and the results of those
calculations will be published elsewhere.

\section{ The pseudo-random numbers}

The study of the logistic map where every successive number is obtained from
the first with the formula
\[
x_{n+1}=rx_{n}(1-x_{n})
\]
constitutes a subject by itself, and the fact that changing the value of $r$
transforms a very simple result in chaos is impressive to everyone who sees
this for the first time. Plotting the result for $\ r$ = 2.8, 3.3, 3.5 etc.,
and finally arriving to the infinite value for the periodicity, about $r=$3.57
provides information about the evolution of the map. When $r=4.0$ the map
generates random numbers distributed in the interval [0,1] as shown in Fig. 1.
\begin{figure}[ptb]
\includegraphics[scale=0.75]{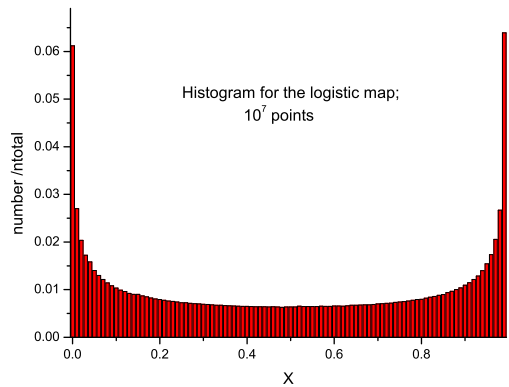}\caption{Histogram showing the
distribution of random numbers in the [0,1] interval, within 100 divisions. }%
\label{F1}%
\end{figure}

The particular distribution can be easily transformed to uniform deviates
using the Ulam-Neumann transformation:
\[
y_{n}=\frac{2}{\pi}\arcsin\sqrt{x_{n}}%
\]

\begin{figure}[ptb]
\includegraphics[scale=0.75]{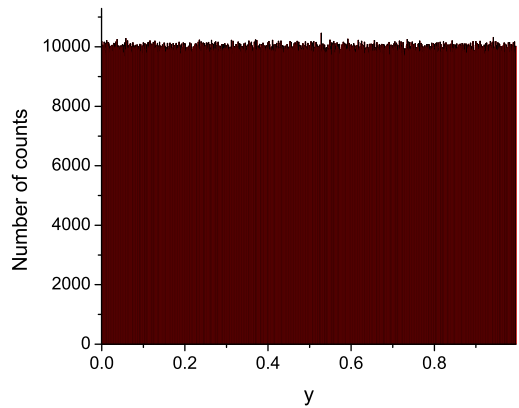}\caption{Result of the Ulam-Neumann
transformation as a histogram with 1000 divisions, to show the noise for
10$^{7}$ numbers}%
\label{F2}%
\end{figure}

Our numbers were generated as follows: A Fortran program with the code to
calculate the formulas above was written, and calculated with a seed $x_{n}$
for $n=0.$ We called this number the \textit{seed }of the sequence. We
discarded the first 300 numbers in every calculation, to avoid any dependence
of it with the initial condition.

The necessary step that follows is to demonstrate that the numbers are
non-periodic, at least within the limits that we are going to use, and beyond.
We check the periodicity with a small program in a PC computer, looking for
any repeated number as the first step, and we proved the numbers are
non-periodic up to 10$^{11}$easily, running the program about 36 hours. As we
wanted to check the periodicity beyond this limit, we used several parallel
and faster computers to check it to 10$^{13}$ numbers. Even limited to
rational numbers, as every digital computer is, we did not observe any period
in the numbers.

The Monte Carlo and other applications require sometimes various independent
sets of random numbers for their development, as in the study of 2D or 3D
Ising models. The question is, if our method allow the generation of
independent sequences. We tried this possibility using different seeds, and we
observed that every sequence obtained, as that is independent of the others,
checking this result as detailed below.

Linear correlation was very easily checked. Even with a small quantity of
numbers as 10$^{7},$ the correlation factor is less than 10$^{-6}.$ Fig 3
shows a graph for a small number of data and two independent sets. Fig 4 shows
the non- correlation of the same two sets between them. Non linear correlation
was also demonstrated using the Hessian matrix.\qquad The mean, variance,
skewness and kurtosis of the distributions were calculated for several seeds,
and the comparison through Student tests and other test demonstrated that the
distributions are equal from the point of view of the parameters, and
different as they are not correlated between them. All these calculations were
extended for a great quantity of numbers, of the order of 10$^{11},$ to make
conclusive the results. It is very easy to obtain distributions of
pseudo-random numbers which are exponential or normal deviates, beginning with
our random deviates and transforming them [7]. Those distributions own the
same properties of non-periodicity as the original one. \begin{figure}[ptb]
\includegraphics[scale=0.75]{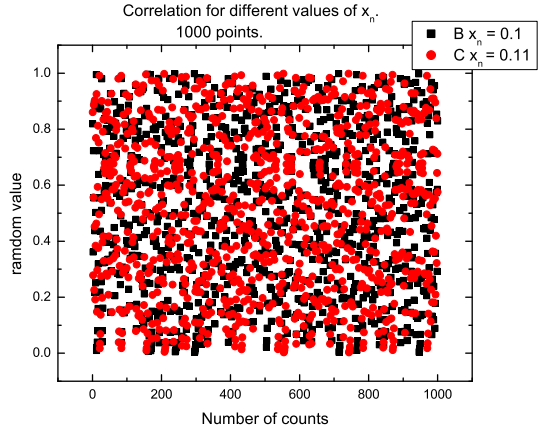}\caption{Graphic to show the
non-correlation between numbers of a set. The plot shows a limited number of
points for clarity (10$^{7}$ points).}%
\label{F3}%
\end{figure}

\begin{figure}[ptb]
\includegraphics[scale=0.75]{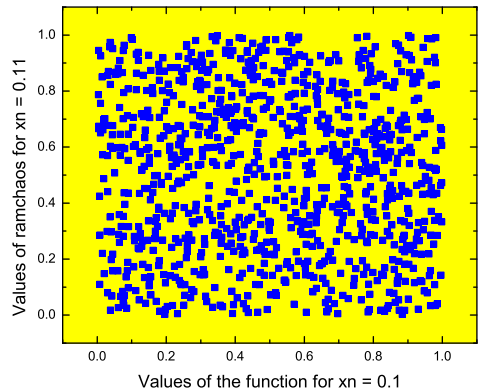}\caption{The same as in Fig 3, but
relating two different sets. The number of points is limited for clarity.}%
\label{F4}%
\end{figure}-

\section{ Applications}

The Ising model has several advantages as a check up for the pseudo-random
generator. One of them is the possibility to calculate it for one, two and
three dimensions using the same desk computer that we were using above. The
second, is that the 2D model was solved exactly, and it is possible to use it
to compare simulations with exact results, including critical parameters, and
delicate properties of the system. Last, but no least, the model is not a
quantum but a classic system when the magnetic field is applied parallel to
the spin orientation direction, and so, it allows a statistical rather than
quantum calculation. With the field perpendicular to the spin orientation, it
becomes maybe the simplest quantum system to be studied. Three models, 1D, 2D
and 3D [6] were calculated, and the meaning of phase transition was
understood. The paradigmatic 2D Ising model was calculated for a 150$\times
$150 lattice, using the simplest method, the single spin-flip and the periodic
boundary conditions using importance sampling, that is, the Metropolis
routine[8]. We began studying a 5$\times5,$ 10$\times10$ and 100$\times100$
lattices, to study the size effect on the calculation, and we found that the
150$\times$150 almost reproduce the exact results of Onsager. The 5$\times$5
and the 10$\times$10 lattices were calculated exactly, using classical
statistical method, in order to compare with the simulations, and those were
simulated using congruential generators, too. When the number of sites in the
lattice increase, this last comparison was impossible, for the generators
overpass the limits of their periodicity. The first calculation was done
within a limited number of iterations for every temperature, 10$^{4}$
iterations, and was afterward repeated with 10$^{5}$ iterations, which was
enough to obtain the magnetization curve shown in fig. 5 and the fluctuations
plotted \ in fig. 6. We used a single routine for all the sets, but obtained
different sets just changing the seed slightly ( from .1 to .11, say). These
curves are but an example of what can be done with out sets of numbers, and we
do intent to demonstrate graphically the efficiency of them. All the
calculations shown here were made using a PC computer, and Fortran \ 77 code,
and for the 2D Ising model, the time necessary to complete the calculation was
less than a day. Most of the - let call them ''serious '' calculations could
require parallel processors, which make faster the calculation, but it is
always possible to do modeling - calculations with a small number of
iterations - with the table computer, which makes the tedious preparation of
the models efficient and comfortable for the researcher.

\begin{figure}[ptb]
\includegraphics[scale=0.75]{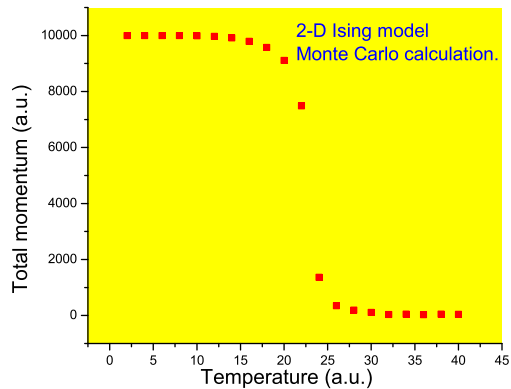}\caption{(Color online)Magnetization
as function of reduced temperature for the 2D Ising model. The points are both
representation of the Onsager theory and the simulation, as the size is small
than the error.}%
\label{F5}%
\end{figure}

As this paper is intended especially to present and analyze the chaotic
sequences generated, we necessarily omit applications on systems as colossal
magnetoresistance compounds, where we simulated the properties of them using
Quantum Monte Carlo and will be published elsewhere, but we made several other
calculations as tests for the numbers. The ''hit or miss'' method of
integration was applied to obtain the number $\pi$ from the integral of a
quarter of a circle, and other integration methods were repeated, both using
our numbers and common congruential pseudo-random numbers [6]. In these cases,
we compared the velocity for the calculations, which is less for the simple
congruential, but faster when the congruential programs are more
sophisticated, as the routine Ran4 in ref. [6]. Following an idea in ref. [1],
we developed a simple and efficient chaotic encripter and decripter. For this
purpose, we used a simplyfied version of the ASCII set, adding and
substracting the numbers generated by the chaotic generator. The code worked
perfectly, and the only information necessary to decript the messages is the
seed. In this sense, we found that the sets generated with very small
differences in the sets are not correlated, and so, a 5 cipher number could be
an excellent seed for this purpose.

Many other calculations were developed to check the pseudo-random numbers.
Just to mention it, we calculated the Ising model using non-periodic and screw
periodic boundary condition for the clusters, identifying size effects in the
calculations; we calculated the same model using cluster flipping methods, and
calculated other systems, as th Heisenberg ferromagnet, using Monte Carlo
methods. The results of those calculations, whenever they conduce to new
results, will be publish elsewhere.

\begin{figure}[ptb]
\includegraphics[scale=0.75]{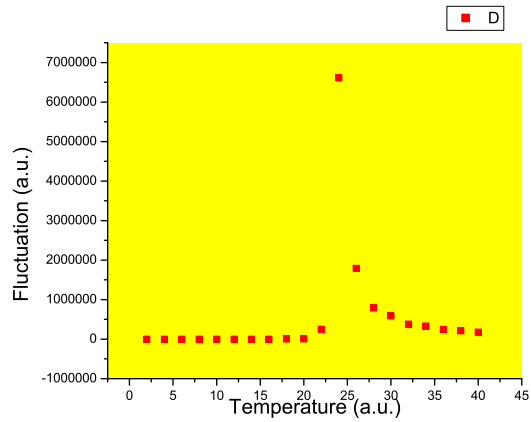}\caption{(color online)Fluctuations in
the 2D Ising model. As in the previous figure, the simulation and the model
coincide. }%
\label{F6}%
\end{figure}\label{}

\section{ Conclusions}

We conclude that we could develop computer programs using the chaotic
properties of the logistic map, for r = 4, where the successive results are
chaotic, deterministic and non-periodic. We studied the sets so obtained,
comparing them with other sets of pseudo-random numbers obtained classically,
and the conclusion is that the chaotic sets accomplish the benefits obtained
with the non-chaotic sets, that is, velocity and determinism, and allow to use
sets as big as necessary, because of their non-periodicity. Most of the
preparation and tests were made in a PC computer, which shows its practicity,
even when calculations with large number of simulations require faster computers.
%\label{}

\end{document}